\documentclass[twocolumn,showpacs,aps,prl,superscriptaddress]{revtex4}

\usepackage{graphicx}
\usepackage{dcolumn}
\usepackage{amsmath}
\usepackage{multirow}

\input pubboard/babarsym

\newcommand{\optbar}[1]{\shortstack{{\tiny (\rule[.4ex]{1em}{.1mm})}
  \\ [-.7ex] $#1$}}
\def\BorBbar    {\kern 0.18em\optbar{\kern -0.18em B}{}\xspace}
\def\DorDbar    {\kern 0.18em\optbar{\kern -0.18em D}{}\xspace}
\def\KorKbar    {\kern 0.18em\optbar{\kern -0.18em K}{}\xspace}

\newcommand{\BABARPubYear}    {03}
\newcommand{\BABARPubNumber}  {036}

\newcommand{\SLACPubNumber} {10271}
\newcommand{\LANLNumber} {00000}

\def\figurebox#1#2#3{%
    \def\arg{#3}%
    \ifx\arg\empty
    {\hfill\vbox{\hsize#2\hrule\hbox to #2{\vrule\hfill\vbox to #1{\hsize#2\vfill}\vrule}\hrule}\hfill}%
    \else
    {\hfill\epsfbox{#3}\hfill}%
    \fi}

\begin{document}


\begin{flushleft}
\babar-PUB-\BABARPubYear/\BABARPubNumber\\
SLAC-PUB-\SLACPubNumber\\
hep-ex/\LANLNumber\\[10mm]
\end{flushleft}

\title{
{\large \bf \boldmath
Measurements of Branching Fractions and \CP-Violating Asymmetries in $B$ Meson
Decays to Charmless Two-Body States Containing a \Kz} 
}

%
\author{B.~Aubert}
\author{R.~Barate}
\author{D.~Boutigny}
\author{F.~Couderc}
\author{J.-M.~Gaillard}
\author{A.~Hicheur}
\author{Y.~Karyotakis}
\author{J.~P.~Lees}
\author{P.~Robbe}
\author{V.~Tisserand}
\author{A.~Zghiche}
\affiliation{Laboratoire de Physique des Particules, F-74941 Annecy-le-Vieux, France }
% ...
\collaboration{The \babar\ Collaboration}
\noaffiliation

\date{\today}

\begin{abstract}
We present measurements of branching fractions and \CP-violating 
asymmetries in decays of $B$ mesons to two-body final states containing a \Kz.
The results are based on a data sample of approximately $88$ million
\upsbb\ decays collected with the \babar\ detector
at the \pep2\ asymmetric-energy $B$ Factory at SLAC.  We measure 
$\BR(\Bp\to\Kz\pip) = (22.3 \pm 1.7 \pm 1.1)\times 10^{-6}$, $\BR(\Bz\to\Kz\piz)
= (11.4\pm 1.7\pm 0.8)\times 10^{-6}$, $\BR(\Bp\to\Kzb\Kp) < 
2.5\times 10^{-6}$, and $\BR(\Bz\to\KzKzb) < 1.8\times 10^{-6}$, where the 
first uncertainty is statistical and the second is systematic, and the upper
limits are at the $90\%$ confidence level.  In addition, the following
\CP-violating asymmetries have been measured:  
${\cal A}_{CP}(\Bp\to\Kz\pip) = -0.05 \pm 0.08 \pm 0.01$ and 
${\cal A}_{CP}(\Bz\to\Kz\piz) = 0.03 \pm 0.36\pm 0.11$.
\end{abstract}


\maketitle

The decays of $B$ mesons into charmless hadronic final states provide important
information for the study of \CP\ violation.  In particular, the study of
the two-body decays $B\to\pi\pi$, $B\to K\pi$, and $B\to KK$ provides 
crucial ingredients for measuring or constraining the values of the 
angles $\alpha$
and $\gamma$, defined by the ratios of 
various elements of the Cabibbo-Kobayashi-Maskawa quark-mixing matrix 
\cite{CKM}:  $\alpha \equiv
{\rm arg}\left[-V^{}_{td}V^*_{tb}/V^{}_{ud}V^*_{ub}\right]$ and
$\gamma \equiv {\rm arg}\left[-V^{}_{ud}V^*_{ub}/V^{}_{cd}V^*_{cb}\right]$. 
In this paper, we present measurements of the branching fractions for $B$ meson
decays to the charmless two-body final states $\Kz\pip$, $\Kzb\Kp$,
$\Kz\piz$, and $\KzKzb$ (unless explicitly stated otherwise, charge conjugate 
decay 
modes are assumed throughout this paper and branching fractions are averaged 
accordingly).  For the $\Bp\to\Kz\pip$ and 
$\Bz\to\Kz\piz$ modes we also report measurements of the direct \CP\ 
asymmetries in the decay rates:
\begin{equation}
{\cal A}_{CP}=\frac{
\Gamma\left(\Bbar\to\bar{f}\right)-\Gamma\left(B\to f\right)}
{\Gamma\left(\Bbar\to\bar{f}\right)+\Gamma\left(B\to f\right)}
\,.
\end{equation}

Measurement of the rates and charge asymmetries for $B\to K\pi$ decays can be
used to establish direct \CP\ violation and to constrain the angle 
$\gamma$ \cite{kpigamma}.
The decay $\Bp\to \Kz\pip$ is dominated by the $b\to s$ penguin process and
in the Standard Model (SM) is expected to have ${\cal A}_{CP}$
close to zero ($<1\%$) \cite{acpkspi}.  Thus, observation of a sizable 
charge asymmetry could be an indication of non-SM contributions to the
penguin loop \cite{acpkspi,newphys}.
The $B \to K\Kb$ decays are characterized by 
penguin and $W$-exchange processes similar to those in 
$\Bz\to\pip\pim$ and can be used \cite{kkburas} to
determine the angle $\alpha$ from the measurement of
the time-dependent asymmetries in $\Bz\to\pip\pim$.  Measurements of the
branching fractions for these decay modes also provide important 
information \cite{kkrosner} regarding rescattering processes.

The measurements presented in this paper are based on data collected
with the \babar\ detector \cite{bbrnim} at the \pep2\ 
asymmetric-energy \epem\ collider \cite{pepii} located at the 
Stanford Linear Accelerator Center.
The sample consists of $87.9\pm 1.0$ million \BB\ pairs produced at the
\FourS\ resonance (``on-resonance''), which corresponds to an integrated
luminosity of about $81\invfb$.
An additional $9\invfb$ of data recorded at an \epem\ 
center-of-mass (CM) energy approximately $40\mev$ below the \FourS\ resonance
(``off-resonance'') are used for background studies.

The \babar\ detector is
described in detail in Ref.~\cite{bbrnim}.  Charged-particle (track) momenta 
are measured in a tracking system consisting of a five-layer, double-sided
silicon vertex detector and a 40-layer drift chamber (DCH), which operate
in a solenoidal magnetic field of $1.5\,{rm T}$.  Particles are identified as
pions or kaons based on the Cherenkov angle measured with a detector of
internally reflected Cherenkov light (DIRC).  The direction and energy
of photons are determined from the energy deposits in a segmented
CsI(Tl) electromagnetic calorimeter (EMC).

Hadronic events are selected on the basis of charged-particle multiplicity and 
event topology.  We reconstruct 
$B$-meson candidates decaying to $\Kz X$, where $X$ refers to
$\pip$, $\piz$, $\Km$ or $\Kzb$.  The $\Kz$ and $\piz$ candidates are reconstructed
in the modes $\Kz\to\KS\to\pip\pim$ and $\piz\to\gamma\gamma$, respectively.
The following selection criteria are applied to the candidate $B$-decay products.

Charged tracks are required to be within the tracking fiducial volume 
and to have at least $12$ DCH hits and a minimum transverse momentum of 
$0.1\gevc$.  Tracks that are not \KS\ decay products are also
required to originate from the interaction point, to be associated with
at least six Cherenkov photons in the DIRC and to have a Cherenkov angle
within $4\,\sigma$ of the expected value for a pion or kaon.

Candidate \KS\ mesons are reconstructed from pairs of 
oppositely charged tracks that form a vertex with 
$\pip\pim$ invariant mass within $3.5\,\sigma$ of the nominal
\KS\ mass and measured proper decaytime greater than five times its
uncertainty.

Candidate \piz\ mesons are formed from pairs of photons having invariant
mass within $3\,\sigma$ of
the nominal \piz\ mass, where the resolution is about $8\mevcc$ for
the candidates of interest.  
Photon
candidates are required to not be matched to a track, to have an
energy of at least $30\mev$, and to have the lateral shower shape expected for
a photon.  The 
\piz\ candidates are then kinematically fit with their mass constrained 
to the nominal \piz\ mass.

The $B$-meson candidate is characterized by two nearly independent
kinematic variables, the energy-substituted mass
$\mes = \sqrt{\left(s/2 + {\mathbf {p}}_i\cdot 
{\mathbf {p}}_B\right)^2/E_i^2- p_B^2}$ and the energy difference
$\Delta E = E^*_B - \sqrt{s}/2$, where the subscripts $i$ and $B$ refer to the
initial $\epem$ system and the $B$ candidate, respectively, the asterisk 
denotes the
\FourS\ rest frame, and $\sqrt{s}$ is the total CM energy.  The pion mass is
assigned to all charged particles in calculating $E^*_B$.  
For $\Bz\to\Kz\Kzb$ and
$\Bz\to\Kz\piz$ candidates, we require $|\Delta E| < 0.11\gev$ and 
$|\Delta E| < 0.15\gev$, respectively.  For $\Bp\to\Kz h^+$ candidates, where
$h$ refers to $\pi$ or $K$, we require $-0.115 < \Delta E < 0.075\gev$.  The
interval is asymmetric in order to select both $\Bp\to\Kz\pip$ and
$\Bp\to\Kzb \Kp$ decays with nearly $100\%$ efficiency.  The $\Delta E$ 
distribution is peaked near zero for the modes with no charged kaons 
and shifted on average $-45\mev$ for $\Bp\to\Kzb \Kp$ decays due to the
pion mass being used for the charged $B$ daughter in the calculation.
The distribution of \mes\ peaks near
the $B$ mass for all modes, and we require $5.20 < \mes < 5.29\gevcc$.

Simulated events \cite{bbrgeant}, off-resonance data, and events in 
on-resonance \mes\ and $\Delta E$ sideband regions are used to study
backgrounds.  The contribution from other $B$-meson decays is found to be
negligible.  The primary background is from random combinations of tracks and
neutral clusters produced in the $\epem \to \qqbar$ events, where $q=u$, $d$,
$s$, or $c$.  In the CM frame,
this background is characterized by its jet structure, in contrast to the
more uniformly distributed decays of the $B$ mesons produced in 
the \FourS\ decays.  
We exploit
this topological difference to suppress such background.  We
require that the angle $\theta_S^*$ between the sphericity axes of the $B$
candidate and of the remaining particles in the event, in
the CM frame, satisfies $|\cos\theta_S^*| < 0.8$.  We also construct a Fisher
discriminant $\cal F$ given by an optimized linear 
combination of $\sum_i p^*_i$ and
$\sum_i p^*_i \cos^2 \theta^*_i$ \cite{alphaprl}, where $p^*_i$ is the
momentum of particle $i$ and $\theta^*_i$ is the angle between its momentum
and the $B$-candidate thrust axis, both calculated in the CM frame.  The shapes
of $\cal F$ for signal and background events are included as probability
density functions (PDFs) in the fits described below.

Signal yields and charge asymmetries are determined from unbinned extended
maximum likelihood fits.  The extended likelihood for a sample of
$N$ $\Kz X$ candidates is
\begin{equation}
{\cal L} = \exp{\left(-\sum_{i}n_i \right)}
\prod_{j=1}^N\left[\sum_i N_i{\cal P}_i(\vec{x}_j;\vec{\alpha}_i)\right],
\end{equation}
where ${\cal P}_i(\vec{x}_j;\vec{\alpha}_i)$ is the probability for a
signal or background category $i$, given by a product of PDFs for the 
measured variables $\vec{x}_j$ of candidate $j$.  
The parameters $\vec{\alpha}_i$ determine the expected distributions of 
measured variables in each category and $n_i$ are the yields determined from
the fit.  We perform separate fits for each of the three samples of $B$
candidates:  $\Bz \to\Kz\piz$, $\Bz \to\Kz\Kzb$, and 
$\Bp \to\Kz h^+$ ($h^+=\pip$ or $\Kp$).  For the two 
neutral $B$ samples there are two categories, signal and background, and
the yield in each category is obtained by maximizing the likelihood.  For these
fits the probability coefficients $N_i$ are the yields (i.e., $N_i=n_i$). 
The charged $B$ decays, $\Bp\to \Kz h^+$, are fit simultaneously with 
two signal categories, $\Bp \to \Kz \pip$ and $\Bp \to \Kzb \Kp$, and two
corresponding background categories.  In addition, the probability
coefficient for each category $i$ is given by 
$N_i = n_i\left(1-q_j{\cal A}_i\right)$, where
$n_i$ is the {\em total\/} yield, summed over charge states, ${\cal A}_i$ is 
the charge asymmetry, and $q_j$ is the measured charge of the given $B$
candidate.  The total yields and charge asymmetries are determined by 
maximizing ${\cal L}$.

\renewcommand{\multirowsetup}{\centering}
\newlength{\LL}\settowidth{\LL}{$8047$}
\begin{table*}[!htb]
\begin{center}
\caption{Summary of results for numbers of selected $\Kz X$ 
candidates $N$, total 
detection efficiencies $\eps$, fitted signal yields $N_S$, statistical
significances $S$, charge-averaged branching fractions $\BR$, charge
asymmetries ${\cal A}_{CP}$, and $90\%$ confidence-level (C.L.) allowed 
asymmetry intervals.  The
efficiencies include the branching fractions for intermediate states
($\Kz\to\KS\to\pip\pim$ and $\piz\to\gamma\gamma$).
Branching fractions are calculated assuming equal rates 
for $\upsbzbz$ and $\Bp\Bm$.
Upper limits for the $\Kzb\Kp$ and $\KzKzb$ branching
fractions correspond to the $90\%$ C.L. and the central values are given
in parentheses.}
\label{tab:results}
\begin{ruledtabular}
\begin{tabular}{lccccccc}
Mode  & $N$ & $\eps$ (\%) & $N_{S}$ & $S(\sigma)$ & \BR($10^{-6}$) & 
${\cal A}_{CP}$ & ${\cal A}_{CP}$ ($90\%$ C.L.)  \\
\hline \\[-3mm]
$\Kz\pip$  & \multirow{2}{\LL}{$8047$} & $13.0 \pm 0.3$ & $255 \pm 20 ^{+11}_{-9}$  &
            $22$  & $22.3 \pm 1.7 \pm 1.1$ & $-0.05\pm 0.08\pm 0.01$ & $[-0.18,0.08]$ \\[2mm]
$\Kzb\Kp$  &                           & $12.8 \pm 0.3$ & $12.4 \pm 8.4 ^{+1.6}_{-2.0}$ &
            $1.7$ & $<2.5 \, (1.1 \pm 0.75 ^{+0.14}_{-0.18})$ &         &  \\[2mm]
$\Kz\piz$  & $2668$                    & $8.6 \pm 0.5$  & $86\pm 13\pm 3$ &
            $12$  & $11.4\pm 1.7\pm 0.8$ & $0.03\pm 0.36\pm 0.11$ & $[-0.59,0.65]$ \\[2mm]
$\KzKzb$   & $754$                     & $8.7 \pm 0.3$ & $4.3^{+5.2}_{-4.1}\pm 1.1$ &
            $1.0$ & $<1.8 \, (0.6^{+0.7}_{-0.5}\pm 0.1)$ & & \\
\end{tabular}
\end{ruledtabular}
\end{center}
\end{table*}

The independent input variables to the fit $\vec{x}_j$ for a given 
event $j$ are
\mes, $\Delta E$, and $\cal F$.  For the fit to the $B^+\to \Kz h^+$ sample we
include the normalized Cherenkov residuals 
$\left( \theta_c - \theta_c^\pi\right)/\sigma_{\theta_c}$ and
$\left( \theta_c - \theta_c^K\right)/\sigma_{\theta_c}$, where $\theta_c$ is
the measured Cherenkov angle of the primary daughter $h^+$, 
$\sigma_{\theta_c}$ is its error and
$\theta_c^\pi\,(\theta_c^K)$ is the expected Cherenkov angle for a pion (kaon).
The quantities $\sigma_{\theta_c}$, $\theta_c^\pi$, and $\theta_c^K$ are 
measured separately for negatively and
positively charged pions and kaons from a control sample of $\Dz\to \Km\pip$ 
originating from $\Dstarp$ decays.

The parameterizations of the PDFs are determined from a combination of
data and simulated events.  The signal \mes\ PDFs for 
$\Bp\to \Kz h^+$ and
$\Bz \to \Kz\Kzb$ are derived from fully reconstructed $\Bp\to\Dzb\pip$ 
decays and are Gaussian.  For $\Bz\to\Kz\piz$, simulated
signal events are employed and the \mes\ PDF is modeled as a Gaussian
distribution with a low-side power-law tail.
We use an empirical threshold
function \cite{argusfcn} to describe the background \mes\ PDFs.  The single
shape parameter of this function is a free parameter in the $\Bp\to \Kz h^+$
and $\Bz \to \Kz \piz$ fits, where the event sample is sufficiently large.  For
the $\Bz \to \Kz\Kzb$ fit this shape parameter is determined from on-resonance
events in $\Delta E$ sidebands.

The $\cal F$
distribution for signal is modeled as a Gaussian function with an asymmetric
width \cite{asymgauss}, where the parameters are determined 
from simulated events.  
For background, it is modeled as a sum of two 
Gaussian functions with 
parameters determined from on-resonance events in \mes\ sidebands.

The signal $\Delta E$ PDFs are derived from simulated events and
are parameterized as a sum of two Gaussian functions for the modes 
$\Bp\to \Kz h^+$ and $\Bz \to \Kz\Kzb$, and as a Gaussian distribution with
a low-side power-law tail for $\Bz\to\Kz\piz$.  The $\Delta E$ distribution
for background is modeled as a second-order polynomial whose parameters
are determined from on-resonance events in \mes\ sidebands.
The normalized Cherenkov-angle residuals are modeled as a sum of two
Gaussian functions.

The results of the maximum likelihood fits are summarized in Table 
\ref{tab:results}.
The $\KzKzb$ final state is an equal admixture
of $\KS\KS$ and $\KL\KL$.  We therefore
assume a $50\%$ probability for the $\KzKzb$ to decay as $\KS\KS$ in computing 
the $\Bz\to\KzKzb$
branching fraction.  We also use the current world averages \cite{PDG2002} for
${\cal B}(\KS\to\pip\pim)$ and ${\cal B}(\piz\to\gamma\gamma)$ in computing
the branching fractions given in Table \ref{tab:results}. 

\begin{figure}[!tbp]
\begin{center}
\includegraphics[width=0.45\linewidth]{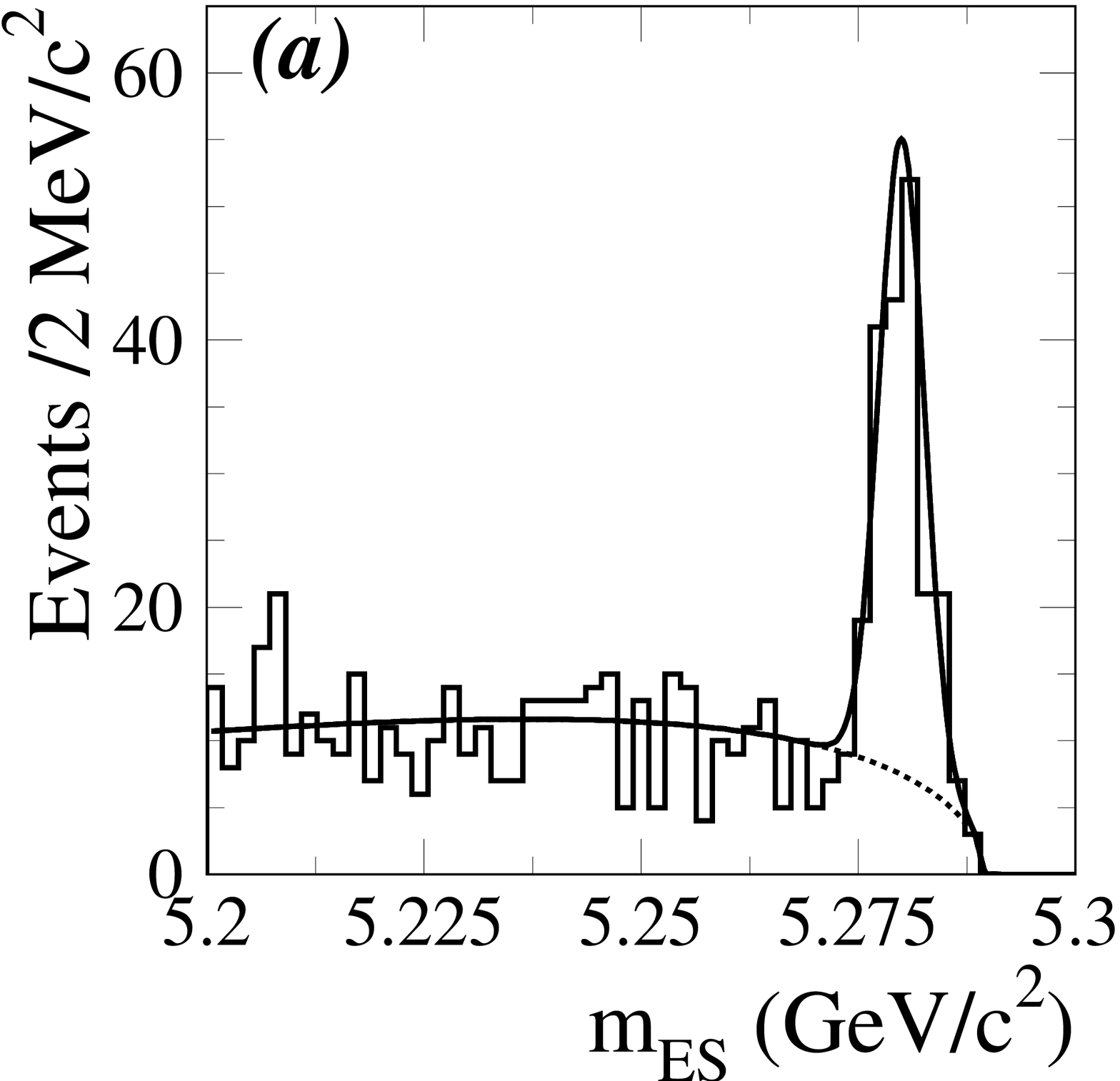}\includegraphics[width=0.45\linewidth]{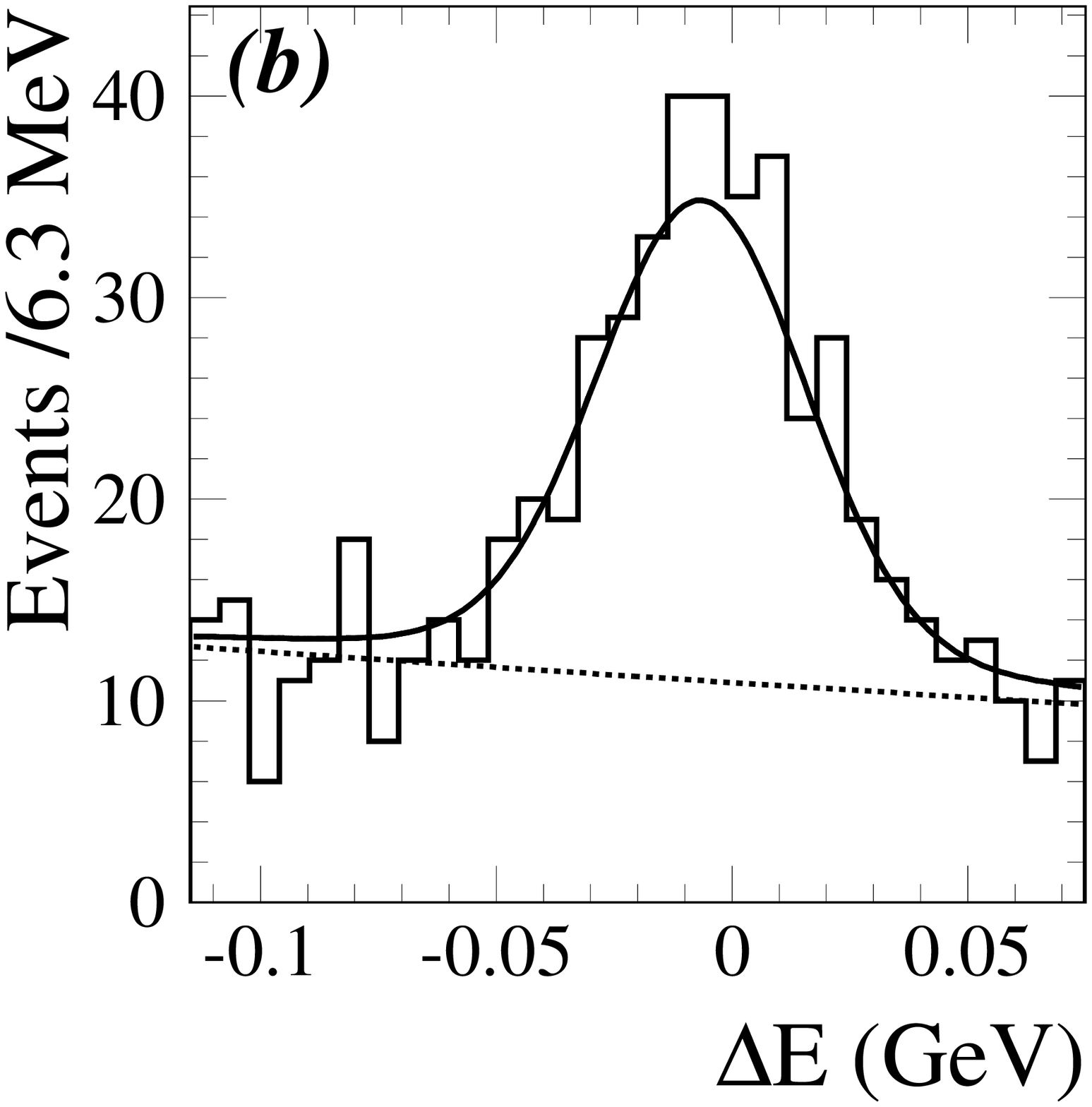}
\includegraphics[width=0.45\linewidth]{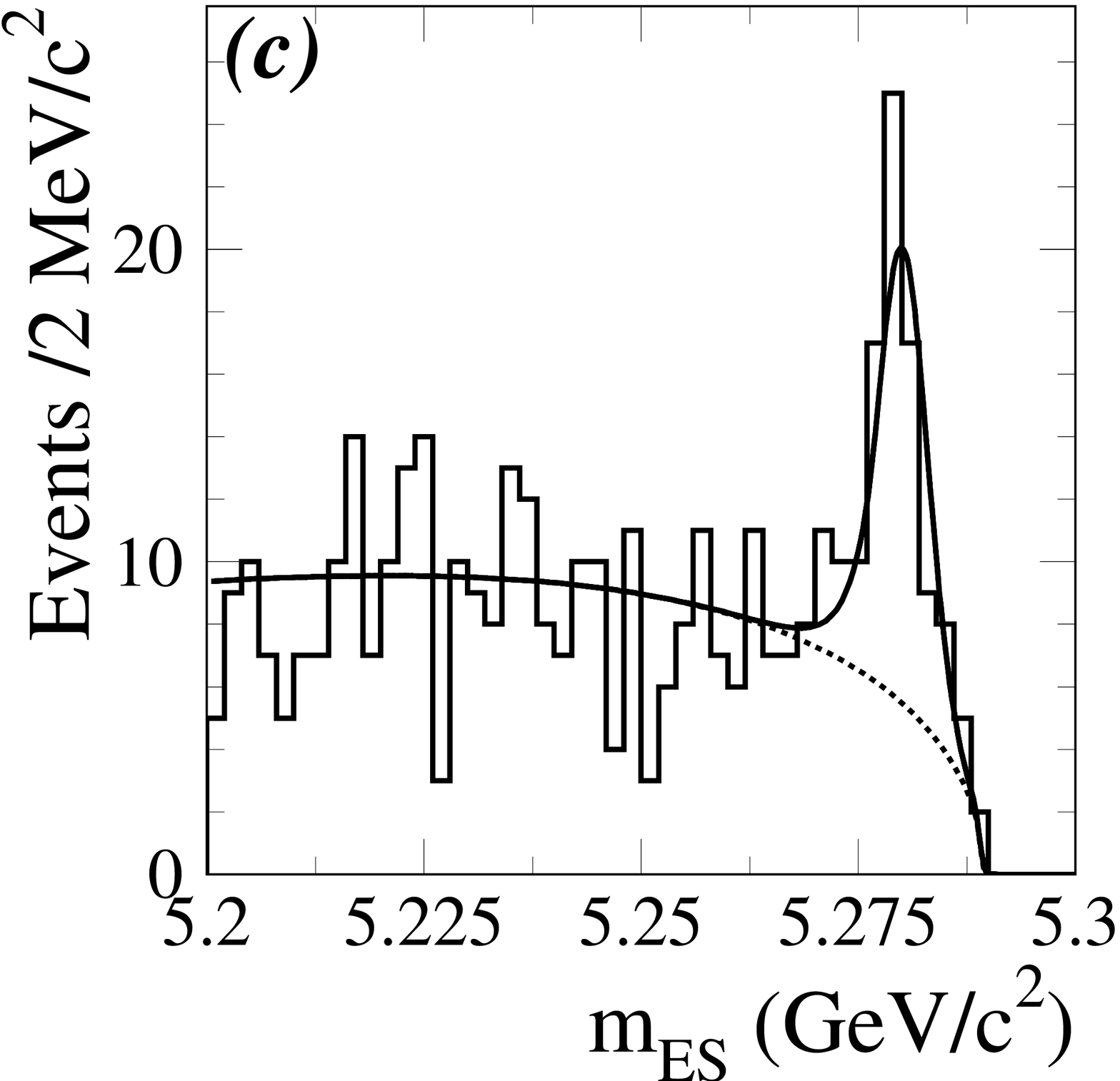}\includegraphics[width=0.45\linewidth]{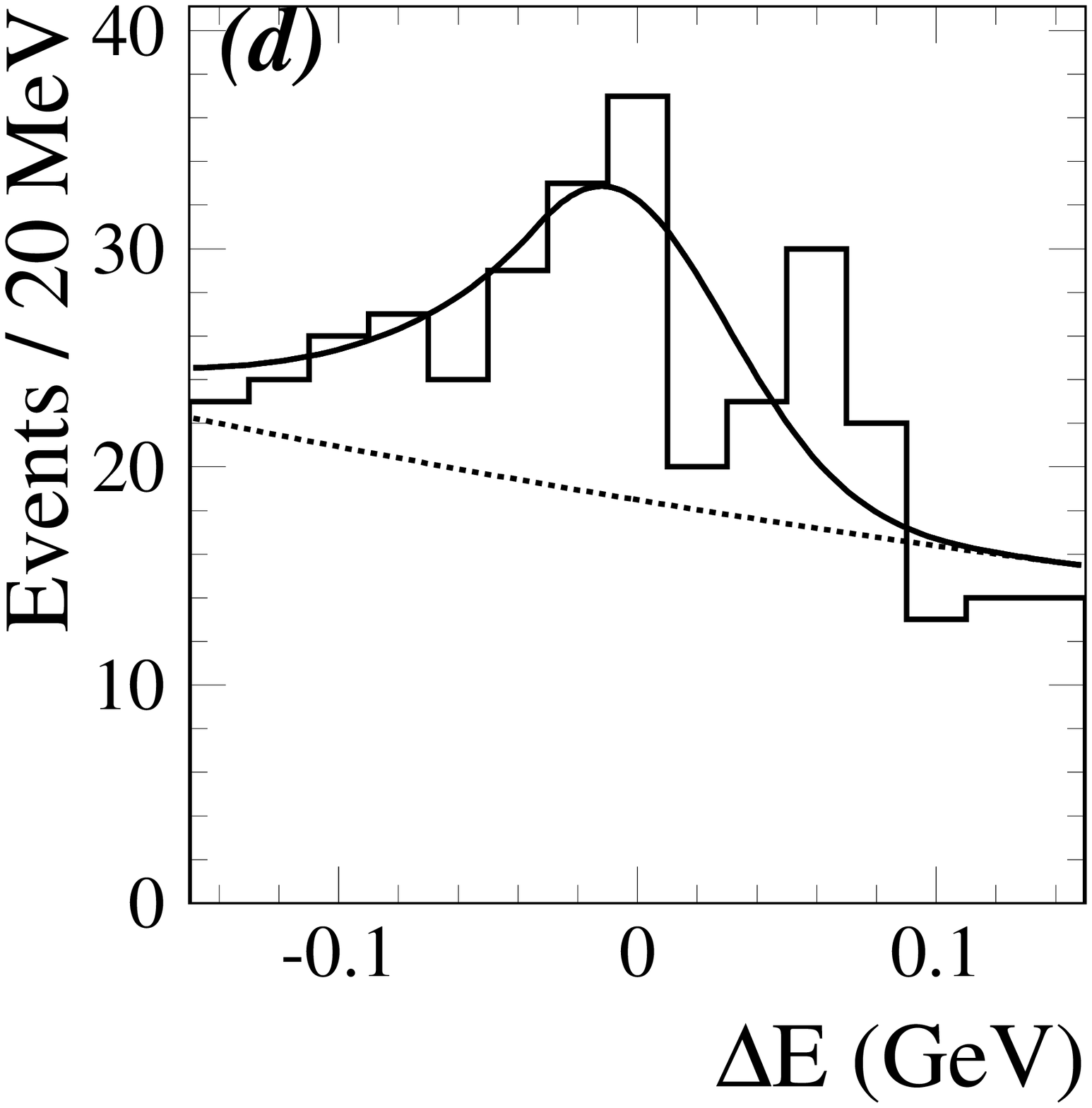}
\caption{Distributions of \mes\ and $\Delta E$ for
(a,b) $\Bp\to \Kz\pip$ and (c,d) $\Bz\to\Kz\piz$ candidates that satisfy an 
optimized requirement on the signal probability, based on all the 
variables except the one being plotted.
The solid curves are projections of the 
fit, while the dashed curves show the background contribution.}
\label{fig:mesde}
\end{center}
\end{figure}

Figure \ref{fig:mesde} shows distributions of \mes\ and
$\Delta E$ for $\Bp\to \Kz\pip$ and $\Bz\to\Kz\piz$ candidates
after selecting on probability ratios to enhance the signal
purity.  The solid curves represent the fit projections after having corrected
for the efficiency of the additional selection.  The efficiencies for these
$\mes\,(\Delta E)$ selection criteria are $70\%\,(93\%)$ and $65\%\,(98\%)$ 
for the $\Kz\pip$ and $\Kz\piz$ states, respectively,
as determined from simulated signal events.

Signal significance is defined as the square root of 
the difference between $-2\ln{\cal L}$ for the best fit and for the 
null-signal hypothesis.  The
upper limit on the signal yield for a given mode $i$ is defined as the value
of $n_i^{\rm ul}$ for which 
$\int_0^{n_i^{\rm ul}}{\cal L}_{\rm max}dn_i/
\int_0^\infty {\cal L}_{\rm max}dn_i = 0.9$, where
$\cal L_{\rm max}$ is the likelihood as a function of $n_i$, maximized with 
respect to the remaining fit parameters.  Branching fraction upper limits are
then calculated by increasing the signal yield upper limit and reducing the
efficiency by their respective systematic uncertainties.

For the $\Bz\to\Kz\piz$ mode, which is a \CP\ eigenstate,
we measure the time-integrated \CP\ asymmetry by
determining whether the {\em other\/} $B$ meson in the event 
decayed as a \Bz\ or
\Bzb\ (flavor tag).  The tagging algorithm is described in 
Ref.~\cite{betaprl}.  The measured asymmetry ${\cal A}_{\rm meas}$ is 
given by ${\cal A}_{CP}/(1+x^2_d)$, where
$x_d=0.755\pm 0.015$ \cite{PDG2002} is the \Bz\ mixing parameter.  The 
dilution of the \CP\ asymmetry by the factor $1/(1+x^2_d)$ is due to the
effect of \Bz-\Bzb\ mixing in the time evolution of the coherent \Bz\Bzb\
system.

Systematic uncertainties in the signal yields arise primarily from imperfect
knowledge of the PDF shapes.  Such systematic errors are evaluated either by
varying the PDF parameters by their measured ($1\,\sigma$) uncertainties
or by substituting alternative PDFs from independent control
samples.  The dominant systematic uncertainty of this type is that associated
with the signal Fisher discriminant for both $\Bp\to \Kz\pip$ ($\pm 7.1$ 
events) and $\Bz\to \Kz\piz$ ($\pm 1.4$ events).  Also contributing to the
systematic uncertainties in the branching fraction measurements are 
the uncertainties in the \KS\ and \piz\ efficiencies, which are
about $3\%$ and $5\%$, respectively.  The systematic uncertainties in the 
charge asymmetries are evaluated by adding in quadrature the contributions
from PDF variations and the upper limit on intrinsic charge bias in the
detector ($\pm 0.01$).  For the measurement of ${\cal A}_{CP}$
in the decay $\Bz\to\Kz\piz$,
there is an additional contribution of $\pm 0.07$ due to 
uncertainties in the tagging efficiencies and mistag fractions.

In summary, we have measured the branching fractions
and \CP-violating charge asymmetries for $\Bp\to\Kz\pip$ and $\Bz\to\Kz\piz$.
No evidence of direct \CP\ violation
has been observed.
We have also searched for the decays $\Bz\to \Kzb\Kp$ and $\Bz\to\KzKzb$ and
set upper limits on their branching fractions at 
$2.5\times 10^{-6}$ and $1.8\times 10^{-6}$, respectively, at the
$90\%$ C.L.
The branching fraction measurements reported here are consistent with
previous measurements of the same quantities 
\cite{cleoprla,hhprl,belleprd}, but have nearly twice the statistical
precision.  Our measured $\Bp\to\Kz\pip$ charge asymmetry is of the same
statistical precision and consistent 
with the value recently reported \cite{belleacp} by the Belle collaboration.
All of the aforementioned results supersede our previous measurements 
\cite{hhprl}, apart from the $\Bz\to\Kz\piz$ charge asymmetry, which has not 
previously been measured.

We are grateful for the excellent luminosity and machine conditions
provided by our \pep2\ colleagues, 
and for the substantial dedicated effort from
the computing organizations that support \babar.
The collaborating institutions wish to thank 
SLAC for its support and kind hospitality. 
This work is supported by
DOE
and NSF (USA),
NSERC (Canada),
IHEP (China),
CEA and
CNRS-IN2P3
(France),
BMBF and DFG
(Germany),
INFN (Italy),
FOM (The Netherlands),
NFR (Norway),
MIST (Russia), and
PPARC (United Kingdom). 
Individuals have received support from the 
A.~P.~Sloan Foundation, 
Research Corporation,
and Alexander von Humboldt Foundation.

\end{document}